\documentclass[sigconf,natbib=true]{acmart}
\usepackage{hyperref}
\usepackage{caption,subcaption}
\usepackage{latexsym}
\usepackage[T1]{fontenc}
\usepackage[utf8]{inputenc}
\usepackage{microtype}
\usepackage{bm}
\usepackage{url}
\usepackage{amsthm}
\usepackage[inline]{enumitem}
\usepackage{multirow}
\usepackage{soul}

\usepackage{xspace}
\usepackage{balance}
\newcommand{\userpmodel}{\textsc{LACE}\xspace}

\AtBeginDocument{%
  \providecommand\BibTeX{{%
    \normalfont B\kern-0.5em{\scshape i\kern-0.25em b}\kern-0.8em\TeX}}}

\copyrightyear{2023}
\acmYear{2023}
\setcopyright{acmlicensed}
\acmConference[SIGIR '23]{Proceedings of the 46th International ACM SIGIR Conference on Research and Development in Information Retrieval}{July 23--27, 2023}{Taipei, Taiwan}
\acmBooktitle{Proceedings of the 46th International ACM SIGIR Conference on Research and Development in Information Retrieval (SIGIR '23), July 23--27, 2023, Taipei, Taiwan}
\acmPrice{15.00}
\acmDOI{10.1145/3539618.3591677}
\acmISBN{978-1-4503-9408-6/23/07}
\begin{document}

\title[Editable User Profiles for Controllable Text Recommendations]{Editable User Profiles for Controllable Text Recommendations}
\author{Sheshera Mysore}
\affiliation{%
 \institution{University of Massachusetts Amherst}
 \country{United States}
}
\email{smysore@cs.umass.edu}

\author{Mahmood Jasim}
\affiliation{%
 \institution{University of Massachusetts Amherst}
 \country{United States}
}
\email{mjasim@cs.umass.edu}

\author{Andrew McCallum}
\affiliation{%
 \institution{University of Massachusetts Amherst}
 \country{United States}
}
\email{mccallum@cs.umass.edu}

\author{Hamed Zamani}
\affiliation{%
 \institution{University of Massachusetts Amherst}
 \country{United States}
}
\email{zamani@cs.umass.edu}

\settopmatter{printacmref=true}

\renewcommand{\shortauthors}{Sheshera Mysore, Mahmood Jasim, Andrew McCallum, \& Hamed Zamani}
\begin{abstract}
    Methods for making high-quality recommendations often rely on learning latent representations from interaction data. These methods, while performant, do not provide ready mechanisms for users to control the recommendation they receive. Our work tackles this problem by proposing LACE, a novel \emph{concept value bottleneck model} for controllable text recommendations. LACE represents each user with a succinct set of human-readable concepts through retrieval given user-interacted documents and learns personalized representations of the concepts based on user documents. This concept based user profile is then leveraged to make recommendations. The design of our model affords control over the recommendations through a number of intuitive interactions with a \emph{transparent user profile}. We first establish the quality of recommendations obtained from LACE in an offline evaluation on three recommendation tasks spanning six datasets in warm-start, cold-start, and zero-shot setups. Next, we validate the controllability of LACE under simulated user interactions. Finally, we implement LACE in an interactive controllable recommender system and conduct a user study to demonstrate that users are able to improve the quality of recommendations they receive through interactions with an editable user profile.
\end{abstract}

\ccsdesc[500]{Information systems~Personalization}
\keywords{interactive recommendation systems; text recommendations; concept bottleneck models; pre-trained language models}

\maketitle

\section{Introduction}
Recommendation systems play a ubiquitous role in influencing the information we consume and the decisions we make. Despite this, these systems fall short of allowing sufficient control to users \cite{eiband2019peoplealgos} or any transparency to system aspects \cite{konstan2021human}. And effective recommenders involve learning opaque user profiles and item representations from user interaction data \cite{zhong2015denseprofile}. The value of control in recommendation has been emphasized by prior work demonstrating greater user satisfaction \cite{jin2017different}, improved trust in the system \cite{jannach2016user}, and an intention to continue consuming content \cite{Yang2019intention}. However, prior work also notes that while users value control, they often prefer hybrid strategies combining automatic methods with interactive strategies for preference elicitation and control, indicating there to be a sweet spot between automation and control~\cite{knijnenburg2011energy, Jin2018musicrec, Yang2019intention}. 

Given the importance of this tradeoff, we develop a performant recommender that facilitates control over recommendations through an editable user profile. For our model, we lay the following goals: (1) to facilitate interactive control, user profiles should be human-readable i.e. transparent, (2) users should be able to edit the profile to express their preferences in various intuitive ways with the recommender system interactively updating its recommendations after profile edits, and (3) the recommender system should make performant recommendations -- ensuring that controllability does not degrade the recommendation quality. Some recent work is relevant to these goals \cite{balog2019transparent, filip2022nlprofiles}. \citet{balog2019transparent} construct transparent user profiles as a set of weighted tags, subsequently used for scrutable recommendation. Despite desirable aspects, their fully transparent approach presents drawbacks in relying on pre-tagged items, not leveraging item content beyond determination of item tags, and remaining inapplicable in the absence of interaction data. On the other hand, \citet{filip2022nlprofiles} make a case for natural language user profiles that may be used to prompt large-language models (LLM) for few-shot scrutable recommendations - while representing an exciting prospect the scrutability of LLMs approaches remains unknown \cite{chen2022iclsen}.

In this paper\footnote{This pre-print extends our \textcolor{blue}{\href{https://dl.acm.org/doi/10.1145/3539618.3591677}{SIGIR 2023 paper}} with extended results in Appendix \ref{sec-extended-results}.}, we introduce a fundamentally different approach for controllable recommendations where our model formulation ensures controllability, effective use of item content, and its use of pre-trained LMs allows effective performance in challenging zero and cold-start scenarios. The starting point for our approach is provided by Concept Bottleneck Models \cite{koh2020cbm} developed for controllable prediction. Our approach, LACE\footnote{Retrieva\ul{l} Enh\ul{a}nced \ul{C}oncept Valu\ul{e} Bottleneck Model}, builds each user profile as a small set of readable concepts retrieved from a large inventory of concepts given user interacted documents. Next, to effectively use user documents, it computes a \emph{personalized concept value} as a function of user documents and the profile concepts. These personalized concept values are computed through a sparse matching of user content to profile concepts computed with an Optimal Transport procedure \cite{peyre2019computational}. These are then used for computing recommendations. LACE admits edits such as positive or negative selections to specify preferences on profile concepts or textual edits to the concepts, which then change the personalized concept values and the recommendations.

We evaluate several aspects of LACE in a series of extensive experiments. We conduct offline evaluations on six real-world datasets spanning three recommendation tasks: scientific paper recommendation, TED Talk recommendation, and paper-reviewer matching for peer review. We validate the efficacy of LACE for generating effective recommendations in three evaluation settings: warm-start, cold-start, and zero-shot. Next, we validate the ability of our model to demonstrate control under \textit{simulated user interactions}. Finally, we implemented LACE in an interactive system and conducted a \textit{user study} to evaluate its interaction ability in a realistic usage scenario. We find LACE to outperform several reasonable baselines in offline evaluations and interactively allow users to make significant improvements to their recommendations. 

Therefore, our contributions include: 1) Proposing a performant model for controllable recommendations, 2) Demonstrating effective empirical performance in numerous evaluation settings, and 3) Establishing the effectiveness of LACE in a realistic user study. Our code is online: \url{https://github.com/iesl/editable_user_profiles-lace}

\section{Problem Formulation}
\label{sec-problem}
\textbf{Desiderata.} In building a controllable recommender we assume access to users $u \in \mathcal{U}$, where each user $u$ has interacted with a set of items $D_u = \{d_{i}\}_{i=1}^{|D_u|}$. Each $d_{i}$ represents a text document of natural language sentences.  To generate recommendations for a user $u$, a ranking system $f$ must generate a ranking $R_u$ over candidate documents $\mathcal{D}$.
\begin{figure}[t]
     \centering
     \fbox{\includegraphics[width=0.4\textwidth]{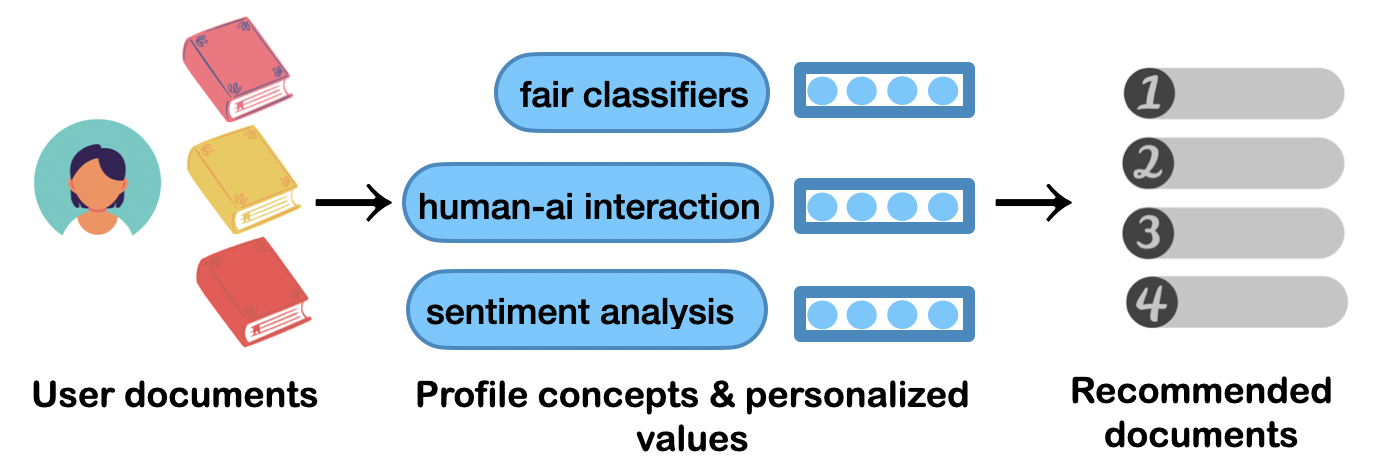}}
     \caption{Our proposed approach, LACE represents users with human readable concept profiles and uses these for controllable recommendations. LACE presents two key novelties: a retrieval-enhanced concept profile allowing users to edit the profile and personalized concept values computed from user documents for performant recommendations.}
     \label{fig-high-level}
     \vspace{-0.5cm}
 \end{figure}
In building an editable user profile, we aim to build a human-readable representation, $\mathcal{P}_u$ of user interactions $D_u$. This is subsequently used for making recommendations as $R_u=f(\mathcal{P}_u, \mathcal{D})$. A user may control $R_u$ by manipulating $\mathcal{P}_u$. This editable user profile $\mathcal{P}_u$ must fulfill the following desiderata:

\ul{D1: Communicate interests to the user.} Users must be able to understand their profile to edit it and control their recommendations. Thus, the profile $\mathcal{P}_u$ should communicate their interests as captured by $D_u$. Importantly, we only aim to have a transparent user profile to facilitate interactive \emph{control} over recommendations. This does not necessitate a ``white-box'' or transparent model \cite{krishnan2020against} -- therefore, we don't pursue this goal.

\ul{D2: Control recommendations via profile edits.} The profile $\mathcal{P}_u$ should provide edit operations to the user, which are then reflected in the recommended results $R'= g(\mathcal{P}'_u, \mathcal{D})$. The profile should broadly support positive and negative preference specifications for interests and correction of errors represented in $\mathcal{P}_u$. Further, to enable users to develop a mental model for control over recommendations, the system must allow fast inference with updated recommendations serving as feedback for user actions \cite{Croft2019interaction, Schnabel2020transparent}. Our goal of control draws on prior work illustrating its benefits \cite{harper2015putting, Yang2019intention}.

\ul{D3: Performant recommendations.} Finally, the profile $\mathcal{P}_u$ should allow for high-quality recommendations before and after profile edits. This follows from users' desire for a sweet spot between automation and control over recommendations \cite{knijnenburg2011energy, Yang2019intention}.

\textbf{Profile Design.}
In this work, we choose to represent $\mathcal{P}_u$ as a set of natural language concepts describing a user's interests. This design follows from a common choice in prior work \cite{kodakateri2009citeseer, bakalov2010introspectiveviews, guesmi2021open} and findings suggesting that users often find concepts/keyphrases to be intuitive descriptors of groups of items \cite{chang2015setprefs, balog2019transparent}.
Specifically, given user documents $D_u$, and a inventory of concepts $\mathcal{K}$, a profile construction model $g$ must induce a user profile of $P$ concepts, $\mathcal{P}_u=\{ k_1, \dots k_P\}$ to describe $D_u$, where $k \in \mathcal{K}$. In our work, interactions include positive or negative selection of concepts in $\mathcal{P}_u$ to indicate interest or disinterest in them or edit actions like adding, removing, or renaming concepts to account for variations or errors in $\mathcal{P}_u$.

\section{Proposed Approach}
\label{sec-proposed-method}
The problem of building a controllable recommender that represents users with human-readable concepts and uses these for making controllable recommendations may be viewed as an instance of a \emph{concept bottleneck model} (CBM) \cite{koh2020cbm, losch2019interpretability}. CBMs are neural network models representing input data $x$ with human-readable concepts $k$ and then using these to predict targets: $x\rightarrow k\rightarrow y$. The concepts allow examination of the model and interventions on $y$.

CBMs involve learning functions $g:x\rightarrow k$ and $f:k\rightarrow y$ from input data paired with concepts and targets: $(x,k,y)$. However, learning a model of this type presents some challenges: 1) Paired user profile data of the form $(D_u, \mathcal{P}_u)$ for training $g$ is often hard to obtain. 2) Since we aim to allow edit actions such as addition, deletion, or renaming of concepts in $\mathcal{P}_u$, $g$ must support these actions and allow interactions to influence downstream predictions. 3) Given that strong text recommendations \cite{bansal2016gru, pappas2013combining} rely on rich user document features, models $g$ and $f$ should leverage neural network features of $D_u$ to generate recommendations.

Our proposed approach (Figure \ref{fig-high-level}), LACE, represents a CBM with two components: \emph{profile construction}: $g:D_u \rightarrow \mathcal{P}_u$, and \emph{ranking} $f:\mathcal{P}_u \rightarrow R_u$. To tackle the challenges outlined above, we present two key novelties in profile construction: i) A retrieval enhanced concept bottleneck: We formulate $g$ as a \emph{retrieval function}, retrieving concepts from a global concept inventory $\mathcal{K}$ to construct a profile $\mathcal{P}_u$ with pre-trained LM encoders. This formulation allows users to make edits to an induced concept bottleneck, with encoders used for concept retrieval also used to encode user edits. Further, use of pre-trained LM encoders allows the construction of $\mathcal{P}_u$ without labeled data $(D_u, \mathcal{P}_u)$. ii) Concept personalization: To leverage features of user documents in representing the profile concepts, each concept is represented with a personalized concept value, $\mathbf{V}_i$ computed as a function of the concept and user documents. Since the personalized concept value is a function of the concept, any user edits to the concept also update the personalized concept value. For this, we leverage Optimal Transport (OT), a method for computing assignments and distances between sets of vectors \cite{peyre2019computational}. Here OT is used to make a sparse assignment of user documents to profile concepts. The assigned document content is then used to compute the personalized concept values. The concepts and their personalized values may also be viewed as keys and values, resulting in a \emph{concept-value bottleneck} model. These personalized concept values $\mathbf{V}^u=\{\mathbf{V_i}\}_{i=1}^{P}$ represent a multi-vector user representation that is used for ranking. For ranking, we follow recent work \cite{mysore2021aspire} and represent candidate documents as multi-vectors $\mathbf{S}^d$ computed from their sentences and generate recommendations by using OT once again to compute distances between sets of user and document vectors: $\mathbf{V}^u$ and $\mathbf{S}^d$. Next, we briefly review Optimal Transport.
\begin{figure}[t]
     \centering
     \fbox{\includegraphics[width=0.4\textwidth]{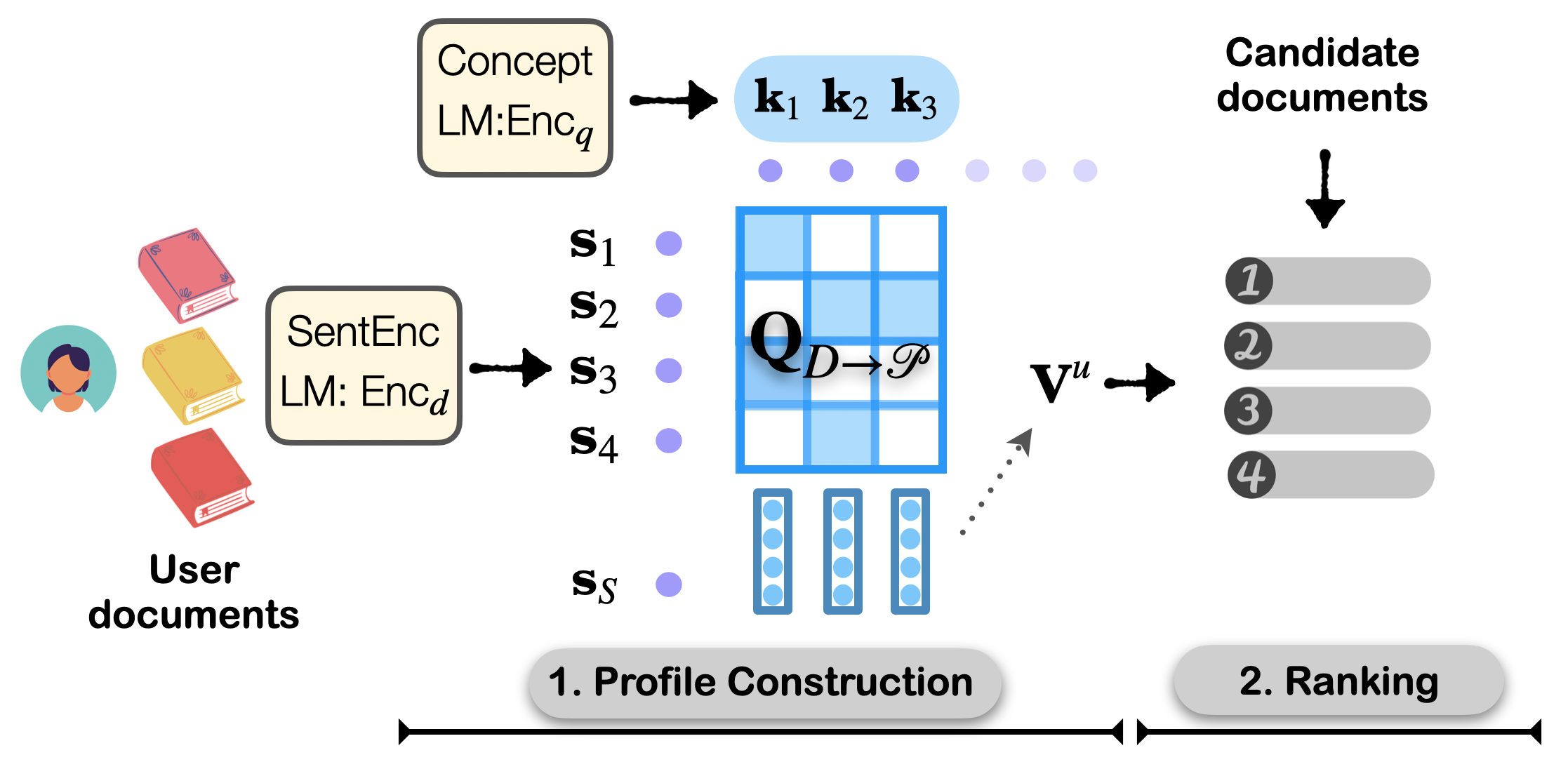}}
     \caption{Profile concepts in LACE, obtained using retrieval, serve as \emph{keys} used to compute \emph{personalized concept values} from  user documents; these are used for ranking.}
     \label{fig-proposed-approach}
     \vspace{-0.5cm}
 \end{figure}

\subsection{Background on Optimal Transport}
\label{sec-optimal-transport}
The optimal transport problem may be seen as a way to compute a minimum cost alignment between sets of points given the pairwise distances between them. Specifically, given the set of points, $\mathbf{S}_p \in \mathbb{R}^{m\times d}$ and $\mathbf{S}_{p^{'}} \in \mathbb{R}^{n\times d}$, and distributions $\mathbf{x}_p$ and $\mathbf{x}_{p'}$ according to which the set of points is distributed. The OT problem involves computation of a soft assignment, the \emph{transport plan} $\mathbf{Q}$, which converts $\mathbf{x}_p$ into 
$\mathbf{x}_{p'}$ by transporting probability mass from $\mathbf{x}_p$ to $\mathbf{x}_{p'}$ while minimizing an aggregate cost $\mathcal{W}$ of moving mass between the points. $\mathcal{W}$ is computed from pairwise costs $\mathbf{C}$. Further, $\mathbf{Q}$ is constrained such that its columns and rows marginalize respectively to $\mathbf{x}_p$ and $\mathbf{x}_{p'}$. 
Therefore, computation of $\mathbf{Q}$ takes the form of a constrained linear optimization problem:
\begin{equation}
    {\mathcal{W}} = \underset{\mathbf{Q'} \in \mathcal{S}}{\texttt{min}}\langle\mathbf{C},\mathbf{Q'}\rangle
    \label{eq-ot-opt}
\end{equation}
Where $\mathcal{W}$ refers to the Wasserstein or Earth Movers Distance and $\mathbf{Q}$ minimizes Eq \eqref{eq-ot-opt}, and the feasible set $\mathcal{S}=\{\mathbf{Q'} \in \mathbb{R}^{m\times n}_{+}|\mathbf{Q'}\mathbf{1}_{n}
=\mathbf{x}_p, \mathbf{Q'}^{T}\mathbf{1}_{m}=\mathbf{x}_{p'}\}$. In our work costs $\mathbf{C}$ are pairwise L2 distances between $\mathbf{S}_p$ and $\mathbf{S}_{p^{'}}$. For computing the solution above, \citet{cuturi2013sinkhorn} introduced an entropy regularized variant of Eq \eqref{eq-ot-opt}. 
This can be used with end-to-end differentiation and allows GPU computation. We leverage this formulation in our work. We do this at two stages of our model: in profile construction, we use the OT plan $\mathbf{Q}$, and for ranking, we use the minimum cost Wasserstein distance $\mathcal{W}$. In our implementations, we use the \texttt{geomloss} package for solving OT problems with entropy regularizer $\lambda=20$ and uniform $\mathbf{x}_{p}, \mathbf{x}_{p'}$

\subsection{Model Description}
\label{sec-model-overview}
\textbf{Retrieval enhanced concept bottleneck.}
Given the user documents $D$ (we drop user subscripts for brevity), we leverage pre-trained language model encoders to retrieve concepts from a concept inventory $\mathcal{K}$ to describe the user documents. This design allows interaction from users who may add concepts not present in the user profile or $\mathcal{K}$, and rename (remove-then-add) existing concepts in $\mathcal{P}$ (Figure \ref{fig-profile-edits}). Therefore the function, $g:D\rightarrow\mathcal{P}$ is factored into a document and concept encoders: $\texttt{Enc}_d$ and $\texttt{Enc}_q$. Specifically, we represent user documents $D=\{s_{i}\}_{i=1}^{S}$ with sentences due to their ability to capture finer-grained information in documents \cite{mysore2021aspire}. Sentence vectors $\mathbf{S}_D$ and concept vectors $\mathbf{K}$, are obtained from $\texttt{Enc}_d$ and $\texttt{Enc}_q$. To construct the profile concepts $\mathcal{P}=\{k_1 \dots k_P\}$ the $\texttt{top-}P$ concepts are retrieved as follows:
\begin{align}
    \mathcal{P} &= \texttt{top-}P(\mathbf{S}_D, \mathbf{K}) \label{eq-profile-topk}\\
    \texttt{dist}(\mathbf{S}_D, \mathbf{k}_i) &= \texttt{min}_{j}\lVert\mathbf{k}_i - \mathbf{s}_j\rVert
    \label{eq-profile-topk-dists}
\end{align}
The distance for individual concepts $\mathbf{k}_i$ is computed as the minimum L2 distance to the sentences. Further, $\mathbf{K}\in\mathbb{R}^{|\mathcal{K}|\times E}$ and $\mathbf{S}_D\in\mathbb{R}^{S\times E}$, where $E$ represents the embedding dimension. This set of retrieved concepts for a user is revised during training as $\mathbf{S}_D$ and $\mathbf{K}$ are updated. In practice, $\mathcal{K}$ may consist of a large number of concepts, so to update $\mathcal{P}$ during training a smaller set of pre-fetched concepts $\mathcal{K}_{f}$ are used for construction of $\mathcal{P}$ in Eq \eqref{eq-profile-topk}.

\textbf{Personalized concept values.} Now given a users profile concepts $\mathcal{P}$ and their embeddings from $\texttt{Enc}_q$ as $\mathbf{K}_{\mathcal{P}}$. The representations of the same concepts across different users will be identical. However, stronger personalization performance can be achieved if user content influences concept representations. To achieve this, we pair each profile concept with a \emph{personalized concept value}: $\mathbf{V}_i$. Specifically, given $\mathbf{K}_{\mathcal{P}}$ and sentence vectors $\mathbf{S}_D$ we compute a soft matching $\mathbf{Q}_{D\rightarrow\mathcal{P}}$ of sentences in $D$ to profile concepts $\mathcal{P}$. We leverage the optimal transport procedure (\S\ref{sec-optimal-transport}) to compute this matching. Computation of $\mathbf{V}_i$ involves computing an assignment weighted average of sentences:
\begin{align}
    \mathbf{V}_{i=\{1\dots P\}} = \frac{1}{\sum_{j=1}^{S}\mathbf{Q}_{ji}}\sum_{j=1}^{S}\mathbf{Q}_{ji}\cdot\mathbf{S}_j
\end{align}
Here we drop subscripts for $\mathbf{Q}$ and $S > P$. Note that computation of $\mathbf{Q}$ involves the use of profile concept vectors $\mathbf{K}_{\mathcal{P}}$ and sentence vectors $\mathbf{S}_D$ through the pairwise cost $\mathbf{C}$ in Eq \eqref{eq-ot-opt}.

These values $\mathbf{V}^u$ represent concept semantics grounded in user content allowing strong personalization performance. Further, OT computes sparse assignments $\mathbf{Q}$ \cite{swanson2020rationalizing}, ensuring that sentences are only assigned to a small number of relevant concepts. Therefore, the concepts partition the sentences into soft clusters described by their concept. This enables users to specify positive or negative preferences for specific concepts, which includes or excludes topical clusters of sentences in generating recommendations. Further, user edits to the text of concepts influence their embeddings, which in turn influence $Q$, $\mathbf{V}^u$, and $R_u$ - allowing edits to reflect in recommendations. In this sense, the profile concepts may be seen as \emph{keys} paired with personalized concept \emph{values}. Finally, since $\mathbf{Q}$ remains differentiable, it allows gradient descent based training.

\textbf{Candidate Ranking.}
Both user and candidate document representations, $\mathbf{V}^{u} \in \mathbb{R}^{P\times E}$ and $\mathbf{S}^{d} \in \mathbb{R}^{L\times E}$ are multi-vector representations. To compute a score $w_{ud}$, to rank the documents $d \in \mathcal{C}$, we seek to align user interests with document content optimally. Further, we expect only a subset of user interests to match the document. Therefore, we score $\mathbf{S}^{d}$ against the top $|\mathcal{T}|$ elements of $\mathbf{V}^{u}$, i.e $\mathbf{V}^{u}_{\mathcal{T}}=\mathbf{V}^{u}[\mathcal{T},:]$ where $|\mathcal{T}| < P$. $\mathcal{T}$ is obtained according to the minimum L2 distances of $\mathbf{V}^{u}[i,:]$ and $\mathbf{S}^{d}$. To compute the distance between multi-vector representations of the user and document, we leverage optimal transport once more \cite{mysore2021aspire}. Having computed pairwise costs $\mathbf{C}_w$ between $\mathbf{V}^{u}_{\mathcal{T}}$ and $\mathbf{S}^{d}$ and a minimum cost alignment $\mathbf{Q}_{w}$, the distance $w_{ud}=\langle\mathbf{C}_w,\mathbf{Q_{w}}\rangle$, is used for ranking.

\textbf{Document Encoder.}
Our approach leverages sentence representation of user documents $D_u$ and candidate items $d \in \mathcal{D}$. Our work uses pre-trained transformer language model encoders for $\texttt{Enc}_d$. Given document $d$, we obtain contextual sentence representations $\mathbf{s}_i \in \mathbf{S}_d$ by averaging contextualized word-piece embeddings from $\texttt{Enc}_d$. In experiments, we use Sentence-Bert \cite{reimers2019sentencebert} for web text and the \textsc{Aspire} for scientific text \cite{mysore2021aspire}. Both models are strong 110M parameter BERT models pre-trained on document-similarity tasks.\footnote{HF Transformers: \url{sentence-transformers/bert-base-nli-mean-tokens}, {allenai/aspire-contextualsentence-multim-compsci}} 

\textbf{Concept Encoder.}
Control over LACE is achieved through profile concepts. These concepts are encoded with $\texttt{Enc}_q$. This requires $\texttt{Enc}_q$ to capture the semantics of concepts to ensure intuitive interaction with users. Further, vectors obtained from $\texttt{Enc}_q$ must be aligned with those obtained from the sentence encoders $\texttt{Enc}_d$. In this sense, $\texttt{Enc}_q$ may be considered a ``query encoder''. For web-text datasets, $\texttt{Enc}_q$ uses Sentence-Bert, making it identical to $\texttt{Enc}_d$. For scientific datasets, we pre-train a concept encoder using the contrastive Inverse Cloze Task (ICT) objective of \citet{lee2019latent} given the lack of performant short text encoders for scientific text. For ICT pre-training, we use 1M concept-document context pairs. Concepts were extracted using the unsupervised concept extraction method of \citet{king2020forecite} from the S2ORC corpus \cite{lo2020s2orc}.

\subsection{Training and Inference}
\label{sec-training-inference}
\textbf{Training.}
Training our model involves fine-tuning the parameters of the sentence and concept encoders: $\texttt{Enc}_d$, $\texttt{Enc}_q$. The primary objective, $\mathcal{L}_{rec}$, updates the parameters of $\texttt{Enc}_d$ from recommendation interactions. Specifically, given a users documents $D_u=\{d_{i}\}_{i=1}^{|D_u|}$, we treat each document $d_{i}$ in turn, as a positive document $d^{+}$ for obtaining candidate sentence vectors $\mathbf{S}^{d^{+}}$ with the other documents $D_{u}^{i}=D_u\setminus\{d_{ui}\}$ used for computing profile values $\mathbf{V}^{u^+}$. Giving us the loss: $\mathcal{L}_{rec} = \sum_{u\in\mathcal{U}}\sum_{i=1}^{|D_u|}\texttt{max}[w^{i}_{ud^+}-w^{i}_{ud^-}+\delta,0]$.
Here, $w^{i}_{uc}$ denotes the distance between the user profile and candidate document. Negative document $d^-$ is randomly sampled from the interactions of a different user $u'$, and margin $\delta=1$. Second, we continue to train $\texttt{Enc}_q$ on an ICT objective \cite{lee2019latent} used to pre-train this encoder to ensure that $\texttt{Enc}_q$ remains updated as $\texttt{Enc}_d$ is trained. Not updating $\texttt{Enc}_q$ with $\mathcal{L}_{rec}$ follows from the intuition that this encoder presents a way for users to interact with the model via edits. Therefore, it only captures the semantics of concepts and concept-sentence matches. Consequently, $\texttt{Enc}_q$ is continually trained on the ICT loss for the user documents $D_u$ using a random half of pre-fetched concepts $\mathcal{K}_f$. Note that though we fine-tune our model, our semi-parametric model with pre-trained encoders allows zero-shot prediction (see \S\ref{sec-rec-eval}). Finally, our ICT objective may be seen as a search objective with \userpmodel trained as a joint search and recommendation model \cite{zamani2020searchrec}. However, we focus on the recommendation task here.

\textbf{Recommendations.}
Making recommendations with LACE involves computing ranked documents $R_u$ from candidate documents $\mathcal{D}$. Since our approach relies on a set of dense vectors to represent users and candidate documents with $\mathbf{V}^{u}$ and $\mathbf{S}^{d}$, these  can be computed as per \S\ref{sec-model-overview} and cached. Ranking involves computation of the Wasserstein distance $w_{ud}$ -- recent work has explored approximate nearest neighbor (ANN) methods for Wasserstein distances \cite{backurs2020OTscalable}. This paves the way for interactive large-scale recommendations with LACE - an important element of our desiderata (\S\ref{sec-problem}, D2). While we indicate the potential of ANN, we leave exploration of this to future work. We opt for a simpler approach - computing $w_{ud}$ for all candidates $\mathcal{D}$ or opting to use LACE as a re-ranker where fast and performant first-stage rankers are available.
\begin{figure}[t]
     \centering
     \fbox{\includegraphics[width=0.43\textwidth]{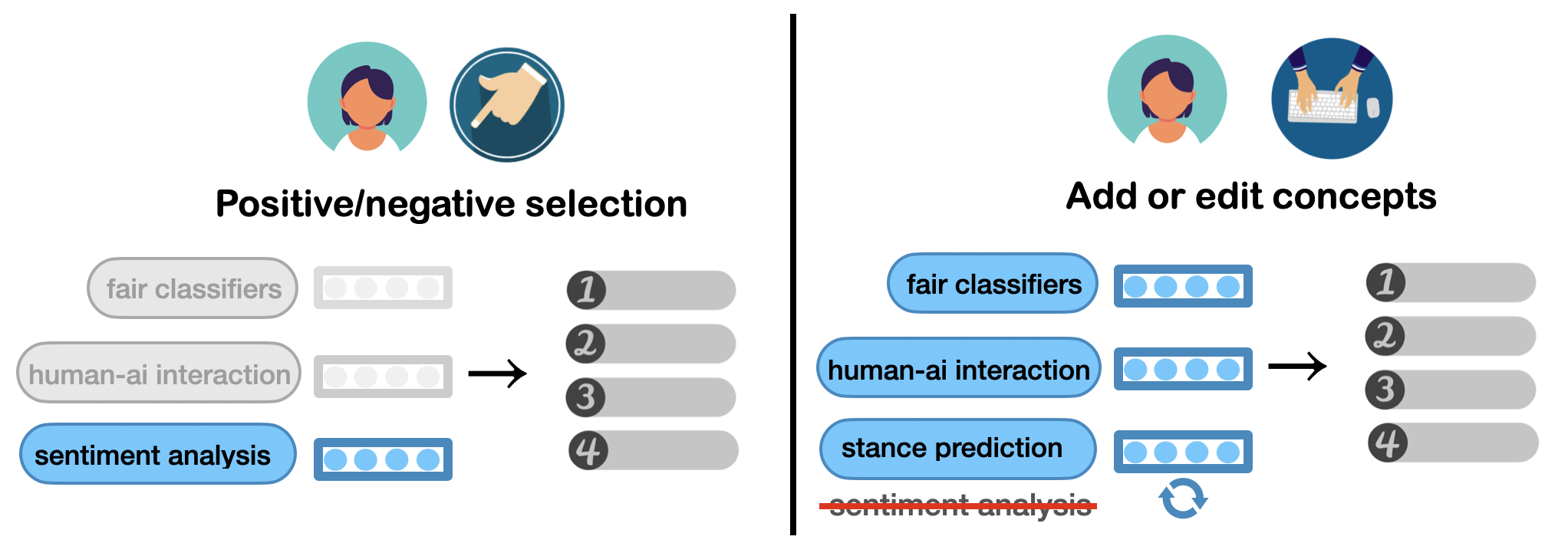}}
     \caption{Users may interact with their concept-based profiles through positive/negative selection of concepts or edit concepts directly through addition, deletion, or renaming.}
     \label{fig-profile-edits}
     \vspace{-0.5cm}
 \end{figure}

\textbf{Interactions.}
Our model allows two forms of interaction through the profile concepts $\mathcal{P}_u$. Users may specify a positive or negative preference for profile concepts or edit the concepts through additions, deletions, and renaming to account for variations and errors (Figure \ref{fig-profile-edits}). These are achieved as follows.
\ul{1. Positive/negative selection.} After construction of profile concepts $\mathcal{P}_u$ and the corresponding values $\mathbf{V}^{u}$, users may positively or negatively select elements of $\mathbf{V}^{u}$ to generate recommendations. E.g., for a user with ``sentiment analysis'' in their profile, positive selection is akin to saying: ``I want only sentiment analysis'' and a negative selection: `I don't want sentiment analysis''. These amount to the recommendations computed from a topical subset of $D_u$. Positive selection results in the positively selected values, $\mathbf{V}^{u}[p,:]$ being used for computing $R_u$. Similarly, negative selection results in a compliment of the selections $\mathbf{V}^{u}[\overline{n},:]$ being used for computing $R_u$. \ul{2. Profile edits.} Users may also directly change the text of concepts in $\mathcal{P}_u$ triggering re-computation of $\mathbf{V}^{u}$ i.e a reorganization of documents in $D_u$. Profile edits may span two types: deletion, addition, or modification of concepts \emph{consistent} with the interests represented in $D_u$, and addition of new interests not captured in $D_u$. While adding new interests is an important interaction \cite{bakalov2010introspectiveviews}, it represents a problem more similar to search, different from recommendation. More generally, we expect profile interactions to take many forms and to be informed by subsequent model results, much like query reformulations \cite{jiang2014longref, chen2021refweb}. A full understanding of this requires thoroughly examining user behaviors \cite{gupta2010socialtags}. However, our user study (\S\ref{sec-userstudy}) explores the efficacy of these interactions for improving recommendations. 

\section{Recommendation Evaluation}
\label{sec-rec-eval}
We conduct an offline evaluation of LACE on three recommendation tasks: scientific paper recommendation, TED talk recommendation, and scientific reviewer-paper matching for peer review. This presents a task where papers suitable and of interest to reviewers must be recommended for review \cite{shah2022peerreview}.

\subsection{Experimental Setup}
\label{sec-exp-setup}
\textbf{Datasets.} For paper recommendation, we use the two public datasets \textsc{CiteULike-A} (Sparsity: 99.78\%) and \textsc{CiteULike-T} (Sparsity: 99.93\%) \cite{Wang2011ctm, Wang2013cult}. Here, user's past history $D_u$ consists of scientific papers (title and abstract) that the user saved in their personal ``libraries'' on the \textsc{CiteULike} platform, collected between 2004-2013. For TED talk recommendation (Sparsity: 99.70\%) \cite{pappas2013combining}, we use a public dataset of users and their saved talks (title and description) which form $D_u$ -- referred to as TEDREC in \S\ref{sec-exp-results}. For reviewer-paper matching, we obtain three private datasets of reviewer-paper assignments from the ICLR 2019, ICLR 2020, and UAI 2019 conferences in collaboration with the OpenReview peer-review platform. Here, users represent expert reviewers, their authored papers (title and abstracts) represent $D_u$, and to-be-reviewed papers represent candidate documents $\mathcal{D}$. The relevance of candidate documents is captured in two ways: bids and assignments. Bids are made by reviewers on a paper to express interest in reviewing the paper, and assignments are made by conference organizers indicating the suitability to review a paper.

\textbf{Implementation Details.}
Next, we describe important dataset-specific and model details and include other hyperparameters in our code release. The concepts in $\mathcal{K}$ used for user profiles must be able to describe documents and be interpretable to users - we use different inventories per dataset. For the TED Talk dataset (web-text), we use the inventory of categories in the dataset, $|\mathcal{K}| = 200$. For Openreview datasets (scientific text), we use user-contributed concepts in the dataset, $|\mathcal{K}| = 8000$. For the CiteULike datasets (scientific text), we extract scientific concepts using the unsupervised method of \citet{king2020forecite} from a corpus of 2.1 million computer science and biomedical papers in the S2ORC corpus \cite{lo2020s2orc} giving $|\mathcal{K}|=116$k. Next, as the encoders in our model are updated during training, we update $\mathcal{P}_u$ from a pre-fetched set of concepts $\mathcal{K}_f$ by retaining 50\% of these for constructing $\mathcal{P}_u$ ($P$ of \S\ref{sec-model-overview}). This results in a variable length profile per user. To build $\mathcal{K}_f$ we retrieve a single concept per sentence from $\mathcal{K}$ for the documents in $D_u$. For web text, we build $\mathcal{K}_f$ with a Sentence-Bert model. For scientific text, we use TFIDF followed by reranking with our pre-trained encoders and retain the top concept. To compute profile-document distances for ranking, we use 20\% of the user profile (i.e.\ $|\mathcal{T}|$ of \S\ref{sec-model-overview}). The choice of $\mathcal{K}_f$ and $\mathcal{T}$ were made from development set performance. 

\textbf{Evaluation Setup.} We evaluate models in warm-start, cold-start, and zero-shot recommendation setups. We evaluate the paper and TED talk recommendation tasks in all three setups and paper reviewer matching in the zero-shot setup since it represents the natural application setup. Specifically,
\begin{enumerate*}
    \item \textit{Warm-start}: Evaluates a method's ability to recommend items seen in training. For every user, we randomly sample 20\% of a user's items and treat these as test items to be retrieved, treating the remaining items \textit{per user} as a training set.
    \item \textit{Cold-start}: Evaluates a method's ability to recommend items unseen in training -- a persistent challenge in recommender systems. Here, a test set is created by randomly sampling 20\% of \textit{all} items in the dataset and ensuring that these are unseen for any user. Models are trained on the remaining 80\% of the data.
    \item \textit{Zero-shot}: This setup is identical to that of cold-start recommendation but is one where no training on interaction data is permitted and represents a task setup explored in more recent work \cite{sileo2022zero}. This is apt in recommendation applications with privacy commitments against use of interaction data as in reviewer-paper matching. More generally reviewer-paper matching presents a natural cold-start setup where candidate items are unobserved during training. It also represents a significant domain shift where authored papers are likely to follow a different distribution than to-be-reviewed papers requiring zero-shot generalization at inference time.
\end{enumerate*}
Model development was performed on a randomly sampled development set of 10\% of users. Following prior work, CiteULike items are chosen such that they are saved by at least 5 users \cite{bansal2016gru, Wang2011ctm}, TEDREC excluded users with fewer than 12 interactions \cite{pappas2013combining}, and OpenReview used all items. 
We report performance in NDCG@20 and recall@20 in the interest of space, noting that result trends hold at rank 5.

\textbf{Baselines.} In the following experiments, we aim to benchmark the recommendation performance of \userpmodel against various classes of established, scalable, and well-performing baselines spanning matrix factorization, content-based, and hybrid models -- often found to outperform more complex architectures \cite{ferrari2019recprogress}. Note, however, that all the baselines cannot be applied in all three evaluation setups.
    \ul{Popular}: A non-personalized baseline recommending items by popularity among users.
    \ul{BPR}: Bayesian Personalized Ranking represents a strong matrix factorization method \cite{rendle2009BPR}. 
    \ul{ALS}: Alternating Least Squares regression represents a matrix factorization method allowing positive interactions to be weighted over negative ones. Popular, BPR, and ALS only work for our warm-start setup.
    \ul{\textsc{Hybrid}}: This method of \citet{bansal2016gru} presents a strong hybrid recommendation method where users are represented by learned latent vectors, and items are represented with a neural network encoder - intended for cold start recommendation. We train this approach with dataset-specific pre-trained transformer language models. We freeze the LM encoder during training. It can only work for our warm- and cold-start setups.
    \ul{\textsc{NeuKNN}}: This represents a transformer LM-based item nearest neighbor method making a recommendation by ranking candidates $\mathcal{D}$ based on the minimum L2 distance to documents in user interactions $D_{u}$. For training, we fine-tune these models for pairwise user document similarity. Zero-shot performance relies only on pre-trained parameters. \textsc{NeuKNN} presents a controllable method for making recommendations by excluding items in user interactions -- in our user study of \S\ref{sec-userstudy-eval}, this serves as a baseline. For \textsc{Hybrid} and \textsc{NeuKNN} models, we use \textsc{specter} \cite{cohan2020specter} for scientific text datasets and Sentence-Bert \cite{reimers2019sentencebert} for TED talk recommendation (\textsc{Hybrid}$_\textsc{Sp/SB}$, $\textsc{NeuKNN}_\textsc{Sp/SB}$). Finally, in warm-start, we use \userpmodel to re-rank the top 100 results for an established approach, ALS -- denoted \userpmodel$_\textsc{rrALS}$.

\subsection{Experimental Results}
\label{sec-exp-results}
Tables \ref{table-rec-warm}, \ref{table-rec-cold}, and \ref{table-rec-or} present our empirical results across datasets and evaluation setups. Here, metrics are in percentage, bold indicates the best metric, and statistical significance of LACE models was measured against all baselines with two-sided t-tests at $p<0.05$ with Bonferroni corrections. Superscripts ($^{\times\#}$) indicate \emph{non} significant results in Table \ref{table-rec-warm}, \ref{table-rec-cold}, and \ref{table-rec-or}. Unmarked results are significant.

\begin{table}[]
\caption{Offline evaluation for warm-start setups.}
\scalebox{0.8}{\begin{tabular}{lllllll}
& \multicolumn{2}{c}{\textsc{CiteULike-A}} & \multicolumn{2}{c}{\textsc{CiteULike-T}} & \multicolumn{2}{c}{TEDREC}\\
\cmidrule(lr){2-3} \cmidrule(lr){4-5} \cmidrule(lr){6-7}
Warm-start       & \small{NDCG@20}   & \small{R@20} & \small{NDCG@20}   & \small{R@20} & \small{NDCG@20}   & \small{R@20} \\\toprule
$^1$Popular         & 0.64 & 1.25  & 1.68 & 3.04 & 7.34 & 12.10\\
$^2$ALS         & 16.82 & 25.55 & 16.58 & 28.58 & 6.49 & 11.07\\
$^3$BPR         &  12.90 & 20.34 & 13.49 & 24.14 & 4.80 & 8.57\\
$^4$\textsc{Hybrid}$_\textsc{Sp/SB}$ & 10.37 & 17.46 & 6.22 & 12.05 & 1.58 & 2.55\\
$^5$\textsc{NeuKNN}$_{\textsc{Sp/SB}}$     & 5.79 & 17.45 & 5.95 & 15.19 & 1.57 & 5.22\\\midrule
\userpmodel       & 8.86 & 20.20$^{\times3}$ &  8.36 & 19.27 & 3.69 & 7.78\\
\userpmodel$_\textsc{rrALS}$       & \textbf{24.86} & \textbf{44.92} &  \textbf{19.74} & \textbf{35.63} & \textbf{12.50} & \textbf{27.27}\\
\bottomrule 
\end{tabular}}
\label{table-rec-warm}
\end{table}
\begin{table}[]
\caption{Offline evaluation for cold-start and zero-shot setups.}
\vspace{-0.3cm}
\scalebox{0.8}{\begin{tabular}{lllllll}
            & \multicolumn{2}{c}{\textsc{CiteULike-A}} & \multicolumn{2}{c}{\textsc{CiteULike-T}} & \multicolumn{2}{c}{\textsc{TEDREC}} \\
\cmidrule(lr){2-3} \cmidrule(lr){4-5} \cmidrule(lr){6-7}
Cold-start            & \small{NDCG@20}   & \small{R@20} & \small{NDCG@20}   & \small{R@20}  & \small{NDCG@20}  & \small{R@20}  \\\toprule
$^1$\textsc{Hybrid}$_\textsc{Sp}$ & 22.93 & 31.64 &   13.61 & 23.45 & 7.32 & 11.51\\
$^2$\textsc{NeuKNN}$_{\textsc{Sp}}$   & 26.31 & 37.36  & 18.38 & 28.75 & 17.49 & 25.69\\
\userpmodel       & \textbf{29.39} & \textbf{39.72} & \textbf{21.20} & \textbf{32.11} & \textbf{17.93}$^{\times2}$ & \textbf{27.06}$^{\times2}$\\\midrule
Zero-shot       &   &  &  & & & \\\toprule
$^1$\textsc{NeuKNN}$_{\textsc{Sp}}$     &  21.25 & 30.46 & 15.21 & 24.07 & 16.30 & 22.77\\
\userpmodel       &  \textbf{27.47} & \textbf{38.86} & \textbf{18.76} & \textbf{29.15} & \textbf{17.41} & \textbf{24.86}\\
\bottomrule
\end{tabular}}
\label{table-rec-cold}
\end{table}

\textbf{Warm-start.} In Table \ref{table-rec-warm}, we first examine the performance of the non-personalized baseline, Popular. In both \textsc{CiteULike-A/T}, it underperforms all other approaches. In TEDREC, however, it sees strong performance -- indicating a strong bias for popular talks in viewer preferences. Next, given the sparsity of these datasets, matrix factorization approaches ALS and BPR show strong performance compared to approaches leveraging content: Hybrid and \textsc{NeuKNN}. This gap is larger in the more sparse CiteULike-T, matching prior understanding \cite{dacrema2019progress}. ALS and BPR also leverage the popularity signal more effectively in TEDREC. Next, while \userpmodel underperforms non-controllable matrix factorization approaches, we see it matches or outperforms approaches leveraging item content. This may be attributed to LACE learning \textit{personalized} representations of user content through its profile values and aggregating the strength of multiple user items in computing user-candidate item similarity. Finally, LACE as a re-ranker for ALS, \userpmodel$_\textsc{rrALS}$, outperforms all other approaches with large percent improvements over best baselines: 47-75\% in \textsc{CiteULike-A}, 19-25\% in \textsc{CiteULike-T}, and 70-125\% in TEDREC. This strategy leverages the benefits of matrix factorization approaches while retaining the benefits of controllability and a content-based model for cold-start and zero-shot recommendation presented by \userpmodel. Re-ranking also presents a path to scaling \userpmodel and adopting it into existing matrix factorization systems.
\begin{table}[]
\caption{Offline evaluation on the Openreview platform in a zero-shot setup -- the realistic setup for this task.}
\vspace{-0.3cm}
\scalebox{0.8}{\begin{tabular}{lllllll}
  & \multicolumn{2}{c}{\textsc{ICLR-2020}} & \multicolumn{2}{c}{\textsc{ICLR-2019}} & \multicolumn{2}{c}{\textsc{UAI-2019}} \\
  \cmidrule(lr){2-3} \cmidrule(lr){4-5} \cmidrule(lr){6-7}
Bids & \small{NDCG@20} & \small{R@20}  & \small{NDCG@20}  & \small{R@20} & \small{NDCG@20}  & \small{R@20}\\\toprule
$^1$\textsc{NeuKNN}$_{\textsc{Sp}}$     & 15.52 & 13.22 & 15.21 & 11.91 & 22.98 & 22.76\\
\userpmodel & \textbf{18.02} & \textbf{15.74} & \textbf{18.33} & \textbf{14.81} & \textbf{28.05} & \textbf{28.17}\\
\midrule
Assignments  & & & & & &\\\toprule
$^1$\textsc{NeuKNN}$_{\textsc{Sp}}$     &  7.52 & 12.85 & 6.51 & 12.58 & 17.21 & 28.16\\
\userpmodel &   \textbf{9.13} & \textbf{15.31} & \textbf{11.88} & \textbf{17.72} & \textbf{20.09}$^{\times1}$ & \textbf{33.13}\\
\bottomrule
\end{tabular}}
\label{table-rec-or}
\end{table}

\textbf{Cold-start.} Table \ref{table-rec-cold} presents results in a setup where test set items are never seen during training. This precludes comparison to Popular, ALS, and BPR. Here, across datasets, a content-based approach \textsc{NeuKNN} sees stronger performance than a Hybrid method indicating the value of content-based representations in the cold-start setup. In the more dense TEDREC dataset, \textsc{NeuKNN}$_\textsc{SB}$ sees stronger performance with \userpmodel matching or slightly outperforming it. This matches prior understanding with item nearest neighbor methods seeing a stronger performance in denser datasets \cite{grvcar2005data}. In the sparser \textsc{CiteULike-A/T} datasets, we see \userpmodel improve upon the best baselines by 5-11\% and 11-15\% respectively. This indicates the potential of \userpmodel for application in the challenging cold-start setup while providing the benefits of interactive control.

\textbf{Zero-shot.} Tables \ref{table-rec-cold} and \ref{table-rec-or} present results in the challenging zero-shot setup where recommendations must be made without any training data -- when presented only with user items $D_u$. Here we only compare against \textsc{NeuKNN}$_{\textsc{Sp/SB}}$ a model which can be applied only with its pre-trained weights. In Table \ref{table-rec-cold}, we see trends similar to the cold-start setup. In \textsc{CiteUlike-A/T}, we see gains of 27-29\% and 21-23\% respectively and 7-9\% in TEDREC. Next, consider Table \ref{table-rec-or}, presenting results on reviewer-paper matching datasets of OpenReview with two measures of relevance: Bids and Assignments. Across datasets, we observe that LACE improves on \textsc{NeuKNN}$_\textsc{Sp}$ by 16-24\% on bids and 17-82\% on assignments. These datasets also represent a difference in distribution between user items (authored papers) and candidate items (to-be-reviewed papers) -- the strong performance of \userpmodel also indicates its robustness to these shifts. Finally, \textsc{NeuKNN}$_\textsc{Sp}$ forms part of the system used for reviewer-paper matching on OpenReview. This performance of \userpmodel also indicates its potential for adoption in the important peer-review application of reviewer-paper matching.

\textbf{Ablation Study.}
In table \ref{table-rec-ablations} we ablate elements of \userpmodel and important baselines to demonstrate the design trade-offs involved. We report performance in zero-shot and cold-start setups on two datasets, given the ability of our models to be applied in these settings and the interest of space.

\ul{Minus CV bottleneck.} Recall that \userpmodel represents documents with sentences and computes personalized concept values, which aggregate sentences for computing user-item scores. A lack of this concept-value bottleneck results in a model which uses sentence vectors of user documents ($\mathbf{S}_D$ of \S\ref{sec-model-overview}) directly for computing user-item scores. In \textsc{CiteULike-A}, we see \userpmodel outperform this model (I) in the zero-shot setup and show smaller gains with training as in cold-start. This indicates the benefit provided by the inductive bias of the concept-value bottleneck, which may be overcome with training data. In TEDREC, we see similar performance for both models, indicating the value of learning dataset-specific patterns. Recall, however, that this simplified model (I) does not offer controllability beyond that offered by \textsc{NeuKNN}.
\begin{table}[]
\caption{Ablations indicating trade-offs in model design.}
\vspace{-0.3cm}
\scalebox{0.85}{\begin{tabular}{rlllll}
  & & \multicolumn{2}{c}{{Cold-start}} & \multicolumn{2}{c}{{Zero-shot}} \\
  \cmidrule(lr){3-4} \cmidrule(lr){5-6}
& \textsc{CiteULike-A}   & \small{NDCG@20}  & \small{R@20} & \small{NDCG@20}  & \small{R@20}\\\toprule
& \userpmodel & 29.39 & 39.72 & 27.47 & 38.86\\
& \textsc{NeuKNN}$_{\textsc{Sp}}$   & 26.31 & 37.36   & 21.25 & 30.46\\
I & LACE\small{$-$CV bottleneck}    & 28.10 & 38.21  & 24.30 & 35.22\\
II & LACE\small{$-$concept values}        &  12.66 & 19.85 & 03.90 & 06.63\\

III & \textsc{NeuKNN}\small{$+$more pretraining} & 26.28 & 37.39 & 24.89 & 35.38\\
\midrule
& \textsc{TEDREC}            & & & & \\\toprule
& \userpmodel & 17.93 & 27.06 & 17.41 & 24.86\\
& \textsc{NeuKNN}$_{\textsc{sb}}$  & 17.49 & 25.69 & 16.30 & 22.77\\
I & LACE\small{$-$CV bottleneck} & 17.78 & 27.02 & 17.85 & 24.79 \\
II & LACE\small{$-$concept values}    & 09.85 & 16.18 & 08.80 & 14.53\\
III & \textsc{NeuKNN}\small{$+$more pretraining} & 17.45 & 25.89 & 16.09 & 21.40\\
\bottomrule
\end{tabular}}
\label{table-rec-ablations}
\end{table}

\ul{Minus concept values.} Next, we consider a model with a concept-based profile as in LACE but not using the personalized concept values of LACE. Instead, this uses embeddings of profile concepts $\mathcal{P}_u$ to compute user-item scores. This approach offers controllability similar to LACE. However, we see (II) this approach consistently underperforms LACE indicating the value provided by the personalized concept values of LACE. Note that this may also be seen as a tag-based recommender as in \citet{balog2019transparent} -- showing personalized concept values to be a more effective use of interacted documents beyond determination of a tags alone.

\ul{More pretraining.} Finally, to examine the benefits of extensive text similarity pre-training for ItemKNN models, we examine the performance of \textsc{NeuKNN} variants with more extensive pre-training.\footnote{HF Transformers, \textsc{CiteULike-A}: \url{allenai/aspire-biencoder-compsci-spec}, TEDREC: \url{sentence-transformers/all-mpnet-base-v2}}. For \textsc{CiteULike-A}, we see (III) improved performance in the zero-shot setup compared to \textsc{NeuKNN}$_\textsc{Sp}$ with gains disappearing upon training as seen in cold-start. In TEDREC, we see more pre-training not to benefit performance, indicating the importance of learning dataset-specific patterns. Broadly, we also see patterns similar to Table \ref{table-rec-cold} and performance comparable or lower than LACE.

\section{Interaction Evaluation}
\label{sec-userstudy}
To evaluate the efficacy of LACE to control recommendations, we first validate some of the interactions provided by LACE through a series of simulations (\S\ref{sec-simumation-eval}). Then we conduct a task-oriented lab evaluation of users interacting with LACE and use the resulting data to evaluate the interactive aspects of LACE (\S\ref{sec-userstudy-eval}).

\subsection{Simulation Evaluation}
\label{sec-simumation-eval}
Before our lab evaluation, we use simulations to evaluate simpler interactions in LACE: validating the positive and negative selection (\S\ref{sec-training-inference}) of individual concepts to perform as expected and synonymous edits (e.g., \ replacing ``passage retrieval'' with ``document retrieval'') to individual profile concepts not to cause large changes in the recommendations received. To measure the efficacy of positive and negative selection, we measure Concept Recall@20 (CR@20) - testing for the presence of a concept in the recommended documents compared to all documents before and after a tuning interaction - a metric also used in prior work \cite{Nema2021bvae}. An effective model should increase CR@20 for positive selection and decrease it for negative selection. We measure robustness through Recall@20 with interactions data, expecting a robust model not to cause large changes after a synonymous edit. Given its cleaner text, we conduct our simulations with user data in \textsc{CiteULike-A}. We begin by selecting the 14 most frequent concepts in user profiles\footnote{Selected to ensure familiarity to the author running the semi-automatic simulation.},
and randomly select 20 users with the concept in their profile. The resulting 280 concept user pairs are used in simulations. For each concept-user pair, we positive/negative select the personalized concept value for the concept in the user's profile, followed by generating recommendations. To simulate synonymous edits, we use a separate Sentence-Bert \cite{reimers2019sentencebert} encoder to encode and generate nearest neighbors for the 14 frequent concepts and select one of the top 20 concepts as a replacement for a user's profile. Table \ref{tab-simulation} depicts the results of our simulations. Positive and negative selection cause CR@20 to increase and decrease significantly (two-sided t-test, p=0.05). We also see synonymous edits not to cause significant changes to R@20. These indicate the effectiveness of simple interactions with LACE.
\begin{table}
\centering
\caption{Simulation evaluation of \userpmodel for positive/negative selection and robustness to synonymous profile edits.}
\vspace{-0.4cm}
\scalebox{0.9}{\begin{tabular}{rrccc}
& & Positive & Negative & Robustness \\
\cmidrule(lr){3-3} \cmidrule(lr){4-4} \cmidrule(lr){5-5}
  &      & \small{CR@20~}$\uparrow$ & \small{CR@20~}$\downarrow$ & \small{R@20~}$=$ \\
 \toprule
\multirow{2}{*}{\userpmodel}  & $^1$Initial & 36.67 & 36.67 & 42.23\\
  & Tuned & 49.75$^{1}$ & 25.94$^{1}$ & 42.14\\
\bottomrule
\end{tabular}}
\label{tab-simulation}
\end{table}

\subsection{Task Oriented User-Study}
\label{sec-userstudy-eval}
Next, we conduct a task-oriented within-subjects lab evaluation of LACE with 20 users consisting of computer science researchers. Here, users interacted with two models \userpmodel (system name, Maple) and \textsc{NeuKNN} (system name, Otter) to receive research paper recommendations, saved the papers they found interesting, and tuned the recommendations to their liking through interactions with the systems. Through the data collected, we aim to answer the following research questions:
\ul{LACE Controllability, RQ1}: Does LACE allow users to improve the recommendations they receive?
\ul{LACE vs. \textsc{NeuKNN}, RQ2}: Does LACE allow users to improve their recommendations more effectively than NeuKNN?
Besides answering these RQs, we also report realistic usage patterns with the data gathered in our user study.

\textbf{System Description.} 
To conduct our study, we developed two interactive recommendation systems for making scientific paper recommendations - Otter and Maple. For both systems, candidate documents $\mathcal{D}$ consisted of 100k computer science papers from the S2ORC corpus \cite{lo2020s2orc}. Of this, 50k were the most highly cited computer science papers to ensure familiarity, and the other half was sampled randomly. Choosing popular items is also common in prior work \cite{balog2019transparent}. Otter used a \textsc{NeuKNN} model and served as our baseline system. Maple used LACE. \textsc{NeuKNN} used a SOTA document bi-encoder model for document similarity \cite{mysore2021aspire} - we denote it as \textsc{NeuKNN}$_\textsc{Asp}$. \userpmodel was implemented similarly to Section \ref{sec-proposed-method}; however, a Sentence-Bert model was used to retrieve profile concepts and optimal transport was implemented using the POT library \cite{flamary2021pot}. Our system used a re-ranking strategy for LACE. Retrieving 500 documents from $\mathcal{D}$ for each $d\in D_u$ using \textsc{NeuKNN}$_\textsc{Asp}$ and re-ranking these documents with \userpmodel. Both systems ran on 2 CPUs and 16 GB RAM. They used identical interfaces, only varying in their recommendation tuning methods.

\textbf{Study Participants.}
Participants for our study were recruited through university mailing lists and announcements made on social media. To ensure that users had formed research interests and could identify papers of interest to them, we asked respondents to confirm that they were involved in a research project and briefly describe their research in our sign-up form. Of the respondents, 20 were selected as participants/users for our study on a first-come-first-serve basis while ensuring they were involved in research. Participants received \$25 gift cards for hour-long participation. All study procedures were approved by the university IRB.
As a proxy for their expertise, participants noted authorship for research papers: 1-5 research papers (14/20), none (5/20), and 6-10 (1/20).

\textbf{User Study Description.}
Our within-subjects study consisted of three main phases, (A) preference elicitation, (B) evaluation of an initial list of recommendations, and finally, (C) multiple iterations of edits to a user profile followed by an examination of the recommended list to improve the initial recommendations. While (A) was performed offline through a web form, (B) and (C) were performed in a 1-hour study session over Zoom.

\ul{Preference Elicitation.} In our offline preference elicitation, users were instructed to submit 2 distinct sets of 4-5 papers or 2 authors of interest to any of their prior, current, or likely future research interests. The submitted information was used to construct a seed set of 20-25 papers representing user documents ($D_u$) per system. To build $D_u$, we expanded the user-submitted papers by randomly sampling the references cited in each submitted paper or by gathering the 20 most recent papers by the submitted author. These methods are common for discovering relevant papers \cite{Athukorala2013compscilit}. Data was gathered through the Semantic Scholar API.

We avoided gathering extensive item ratings to keep a low burden on participants. Further, having participants submit papers of interest to their research ensured that they were experts in these areas and could evaluate recommendations relevant to their research interests. Additionally, a semi-automatic method for gathering $D_u$ also ensured that while many papers in $D_u$ were topically relevant to users, some were undesirable - necessitating tuning. Finally, our study also used distinct sets of seed papers $D_u$ per system; this ensured that users remained engaged in examining recommendations from both systems. However, this meant that comparisons between both systems could only be made in aggregate across all users.

\ul{Study Procedure.} Next, in an hour-long session conducted via Zoom, users used the Maple and Otter systems for 20-25 minutes each or until they decided to stop. They did not know the proposed vs baseline systems, and the order of system use was random to prevent fatigue or learning effects privileging a system. 

In Maple, users first skimmed the seed papers $D_u$ to familiarize themselves with their contents. Next, they examined the inferred profile concepts $\mathcal{P}_u$. Here, users removed concepts if they were redundant, nonsensical, or did not represent the seed papers or added concepts if there were aspects of the seed papers which were missing in $\mathcal{P}_u$. These were used to compute personalized concept values $\mathbf{V}^u$ and produce an initial list of 30 recommendations $R^0_u$. Users were told to examine the recommended papers and save the ones they wanted to read in more detail. To ground their interest and mirror preference elicitation, users were encouraged to consider papers relevant to prior, current, or future research interests. Following examination of $R^0_u$, users were free to find as many more interesting papers as they could by interacting with the profile, examining the recommendations, and saving the ones they found interesting. Users could make positive or negative selections from $\mathcal{P}_u$ to refine their recommendations by focusing on specific concepts or excluding some concepts, respectively. Alternatively, they could edit concepts in $\mathcal{P}_u$ if there were aspects of $D_u$ they wanted to focus on or to accomplish other intents. 

The study procedure for Otter mirrored that of Maple. However, users did not make concept corrections in Otter (\textsc{NeuKNN}). Further, to refine their recommendations, users could only exclude papers in $D_u$ and recompute their recommendations (i.e negative/positive selection) -- the only interaction possible with \textsc{NeuKNN}. 
We gathered tuning actions (additions, deletions of concepts or seed items), recommended lists, and saved papers in both systems. In our subsequent analysis, user saves are treated as binary relevance measures.

\textbf{User-study Results.}
Our user study allows us to answer RQ1 and RQ2, examining if \userpmodel allows users to improve their recommendations and if it is more effective than a baseline \textsc{NeuKNN}. We answer both questions through ranking metrics: MRR, NDCG@5, NDCG@20 in Table \ref{tab-ustudy-ranking}. We omit recall metrics since they cannot be computed from our study, and superscripts indicate statistical significance at p=0.05 with a paired t-test. Note that, Table \ref{tab-ustudy-ranking} shows higher metrics than \S\ref{sec-exp-results} due to the fully judged recommendation lists obtained from users in the user study, unlike the incomplete relevance labels of implicit feedback datasets.
\begin{table}
\centering 
\caption{User study evaluation of \userpmodel compared against the baseline \textsc{NeuKNN}$_\textsc{Asp}$ for improving recommendations.}
\scalebox{0.9}{\begin{tabular}{rrlll}
 &      & \small{MRR~} & \small{NDCG@5~} & \small{NDCG@20~}\\
\toprule
 \multirow{3}{*}{\textsc{NeuKNN}$_\textsc{Asp}$} & $^1$Initial & 66.97 & 48.30 & 69.16\\
  & Tuned    & 85.38 & 67.54$^{1}$ & 82.03$^{1}$\\
 & + Gain  & 18.42 & 19.24 & 12.88\\
\midrule
\multirow{3}{*}{\userpmodel} & $^a$Initial & 70.76 & 50.93 & 71.68\\
  & Tuned    & 90.00$^{a}$ & 74.83$^{a}$ & 86.16$^{a}$\\
  & + Gain  & 19.24 & 23.90 & 14.48\\
\bottomrule
\end{tabular}}
\label{tab-ustudy-ranking}
\end{table}

\ul{LACE Controllability.} Comparing Initial vs. Tuned performance in Table \ref{tab-ustudy-ranking}, users saw statistically significant gains of 20-47\%  through interactions with \userpmodel. Therefore we answer RQ1 in the affirmative - \userpmodel is effectively able to improve the quality of recommendations users receive. We also note that \textsc{NeuKNN}$_\textsc{Asp}$ also saw gains of 18-40\% from tuning. In using \userpmodel and \textsc{NeuKNN}$_\textsc{Asp}$, users made $2.65$ and $2.20$ tuning iterations respectively. Both systems were used for the same duration or until users choose to stop.

\ul{LACE vs \textsc{NeuKNN}.} To compare the two systems, we examine both models' Initial, Tuned, and Gain performance. In Table \ref{tab-ustudy-ranking}, we note that \userpmodel outperforms \textsc{NeuKNN}$_\textsc{Asp}$ at the Initial and Tuned stages by 4-6\% and 5-10\%, respectively. We also note slightly larger Gained metrics for \userpmodel. However, these were not statistically significant. Therefore, our results indicate that \userpmodel sees improved performance before and after tuning compared to \textsc{NeuKNN}$_\textsc{Asp}$. However, larger-scale studies with diversity in users, and larger candidate document sets are necessary to establish significance.

\ul{Characterizing Usage of LACE.} Next, we characterize usage with \userpmodel as implemented in Maple. Based on the changes users made to $\mathcal{P}_u$ for redundancy and correctness we found the initially inferred $\mathcal{P}_u$ to have a precision of $74.25\%$ on average. Here, users deleted 5.15 concepts and added 1.25 concepts. Fewer additions indicate that the inferred profiles had sufficient coverage of $D_u$, but suffered from redundant or incorrect concepts. In tuning their recommendations, 
users made $7.2$ positive/negative selections and $0.68$ additions to their initially corrected profile, indicating a preference for selection operations over edits to $\mathcal{P}_u$.

\section{Related Work}
Next, we discuss the rich body of prior work on which we build.

\textbf{Interacting with Recommenders.} A line of work has explored interfaces to control recommenders and their influence on users. This line of work has explored user profile-based interaction, with profiles constructed automatically \cite{Musto2020umap, guesmi2021open, bakalov2010introspectiveviews} or from user input \cite{rahdari2021studentpro}. Here, automatic methods often construct profiles using supervised methods for keyphrase or entity extraction, enrich the extractions via linking to existing knowledge bases, and then, use them to compute recommendations. This line of work has also explored other forms of control, ranging from a selection between different algorithms, applying keyword filters to a generated list of recommendations \cite{knijnenburg2011energy, Jin2018musicrec, harambam2019controlui}, and changes to algorithms themselves \cite{harper2015putting}. Different from our work which contributes a performant interactive recommender this work has used existing methods and focused on studying the rich ways in which recommenders influence users.

\textbf{Critiquable Recommenders.} Critiquable recommenders allow control over recommendations in one of three ways, at present: 1) \emph{one-time user feedback} on recommended item explanations followed by \textit{retraining} whole or parts of a model \cite{lee2020limeade, Ghazimatin2021elixir}. 2) \emph{Conversational feedback} via item keyphrases or explanations with a latent user representation updated incrementally given a critique \cite{luo2020latentlinear,expl2021antognini, antognini2021fast, yang2021bayesiancrit, li2020rankopt, wu2019langcrit}. 3) \emph{Feedback via item attributes} which endow latent user or item dimensions with information of item attributes \cite{Nema2021bvae, wang2021controllable}. This allows user feedback to influence user or item representations, which are then reflected in recommendations. Each of these bears differences to \userpmodel. Our approach does not require retraining as in one-time user feedback methods. 
Next, while methods for one-time or conversational feedback via explanations control recommendations by interaction with \emph{individual items}, \userpmodel allows control over sets of items, a more intuitive and efficient structure for expressing preferences \cite{chang2015setprefs, balog2019transparent}. Finally, current methods for conversational and item attribute feedback rely on observing keyphrase usage of users or item attributes for training, not allowing expansion at test time -- \userpmodel allows test expansion of keyphrases/attributes to user-specified keyphrases, offering greater flexibility. 

Another notable aspect of the work in critiquable recommendations is their use of variational autoencoders to capture user preferences via a single latent vector and model components to update this vector from feedback \cite{luo2020deepcrit, antognini2021fast, yang2021bayesiancrit, Nema2021bvae, wang2021controllable}. The concept-value bottleneck of \userpmodel models \textit{multiple} user interests explicitly as human readable concepts which can be directly interacted with -- not requiring modeling for aligning latent dimensions to human interpretable ones or updating them with feedback.

\textbf{Richer User and Item Modeling.} As in \userpmodel, prior work in information filtering and recommendation has developed content-based recommenders \cite{bansal2016gru, Wang2011ctm, lops2011content} allowing performant recommendation in cold-start and zero-shot setups. Further, prior work has also modeled the multiple interests of users in collaborative filtering models via multiple \emph{latent} prototypes\cite{weston2013multiinterest, barkan2021anchorcf, yang2021localf}. Our work primarily differs from these in building interactive and transparent user profiles to control content-based recommenders.

\textbf{User Profile Construction.} A line of work has sought to build user profiles for a range of applications leveraging approaches in matrix factorization \cite{cao2017knownfor}, learning to rank \cite{ribeiro2015tagexpert}, and information extraction \cite{li2014weaklysupprofile, ching2008folksononies}. This line of work often attempts to build general-purpose user profiles while leveraging labeled data such as tagging behavior of users \cite{cao2017knownfor}, profile attribute values extracted from social networks \cite{li2014weaklysupprofile}, and user-assigned document tags \cite{ching2008folksononies, ribeiro2015tagexpert}.
While this line of work leverages supervised data for constructing profiles we contribute a method for inducing user profiles in the absence of labeled data \emph{and} influencing a downstream recommender.


\section{Conclusion and Future Work}
Our paper introduces a novel retrieval-enhanced concept-value bottleneck model, LACE, for constructing a human editable user profile and making performant text recommendations. 
We demonstrate strong performance in 3 recommendation tasks and 6 datasets in offline evaluations. Then we validate the controllability of LACE through simulated edits and a task-oriented user study. We demonstrate that users can make significant improvements to their recommendations through interaction with LACE.

LACE presents several opportunities for future work. The concept-value user representation may be used for controllable personalization in other applications, e.g. search \cite{taveen2005ps} and text generation \cite{dudy2021refocusing, salemi2023lamp}, perhaps by augmenting transformer language models with a rich and compact personalized memory \cite{wu2022memorizing}. Further, richer structured user data in the form of personal knowledge graphs may motivate more structured profile representations and accompanying interactive learning and inference algorithms \cite{balog2019pkg}. Further, the intuitive profile edit interactions supported in LACE call for the design of interactive recommenders leveraging this strength. These may then be studied and evaluated in larger-scale online evaluation spanning impactful applications, such as peer-review and text recommendation -- presenting benefits to end users and providing a rich canvas for future research spanning multiple communities.

\medskip
\begin{acks}
We thank anonymous reviewers for their invaluable feedback. This work was partly supported by the Center for Intelligent Information Retrieval, NSF grants IIS-1922090 and 2143434, the Office of Naval Research contract number N000142212688, and the Chan Zuckerberg Initiative under the project Scientific Knowledge Base Construction. Any opinions, findings and conclusions or recommendations expressed in this material are those of the authors and do not necessarily reflect those of the sponsors.
\end{acks}

\appendix
\section{Extended Results}
\label{sec-extended-results}
In Section \ref{sec-exp-results} we evaluated \userpmodel on private datasets obtained from the Openreview platform for reviewer-paper matching, here, we report additional results on the recently released open dataset of \citet{stelmakh2023gold} for this task. We refer to the dataset as \textsc{RAPRatings}. This presents a zero-shot task where models must rank candidate documents per reviewer (i.e.\ user) based on reviewers prior publications with no training on reviewer-candidate ratings. Table \ref{tab-rapratings-ranking} and \ref{tab-rapratings-clf} present these results.
\begin{table}[t]
\caption{\userpmodel compared to baseline models reported in \citet{stelmakh2023gold}. These models represent commonly used reviewer-paper matching models in computer science conferences. A lower Loss indicates greater similarity between model-predicted rankings and user rating-induced rankings.}
\begin{tabular}{rcc}
\textsc{RAPRatings} & Loss & 95\% CI for Loss\\
\toprule
TPMS & 0.28 & [0.23, 0.33]\\
TPMS$_{\text{FullText}}$ & 0.24 & [0.19, 0.30]\\
ELMo & 0.34 & [0.29, 0.40]\\
ACL & 0.30 & [0.25, 0.35]\\
Specter & 0.27 & [0.21, 0.34]\\
Specter+MFR & 0.24 & [0.18, 0.30]\\
\midrule
LACE$_{|D_u|=20}$ & 0.23 & [0.18; 0.29]\\
LACE$_{|D_u|=\text{All}}$ & \textbf{0.22} & [0.17; 0.28]\\
\bottomrule
\end{tabular}
\label{tab-rapratings-ranking}
\end{table}

To build \textsc{RAPRatings}, Stelmakh et al.\ solicit recently read research papers from 58 computer science researchers, these form candidate papers per reviewer. Then, reviewers are instructed to provide ratings (1-5), of their expertise for reviewing each candidate paper in a peer-review setup. Their authored papers constitute a user profile, $D_u$. Stelmakh et al.\ report system performance through a metric similar to Kendall's Tau (Loss in Table \ref{tab-rapratings-ranking}), measuring rank correlation between system-predicted rankings and rankings induced by user ratings. Additionally, user ratings are used to construct two classification tasks on pairs of papers per user. An ``Easy'' classification task checks the system's ability to score a highly-rated document higher than a low-rated document (e.g.\ papers rated 1 vs 5). A ``Hard'' classification task checks the system's ability to score highly-rated document higher than slightly lower rated document (e.g.\ papers rated 4 and 5). These are reported as EA and HA in Table \ref{tab-rapratings-clf}. Finally, the baselines of Stelmakh et al.\ represent methods similar to \textsc{NeuKNN}, of Section \ref{sec-exp-setup}, with sparse (TPMS, TPMS$\text{FullText}$) and dense document representations (ELMo, ACL, Specter, Specter+MFR). These represent models commonly used at leading computer science conference venues for reviewer-paper matching. The implementation details for \userpmodel resemble those used for CiteULike in Section \ref{sec-exp-setup}.

In Table \ref{tab-rapratings-ranking}, we report performance for \userpmodel with $D_u$ set to the 20 most recently authored papers for a user (following Stelmakh et al.) and using all user papers ($|D_u|=\text{All}$). In both settings, we see \userpmodel outperform prior methods. Further, we note a slight improvement in \userpmodel from using all authored papers instead of the most recent 20. Next, in Table \ref{tab-rapratings-clf}, we see \userpmodel outperforms several baseline methods on EA. However, \userpmodel underperforms several baselines on HA -- with TPMS$_{\text{FullText}}$ outperforming other methods. This indicates that while \userpmodel may ensure that many reviewers are assigned relevant papers, methods modeling very fine-grained similarity are likely to assign the best possible paper for a reviewer. Note, however, that this may come at the cost of assigning irrelevant papers to others. We defer further analysis of such behaviors to future work -- noting that a joint treatment of reviewer-paper similarity methods and their use in downstream assignment algorithms \cite{stelmakh19prfa, kobren2019matching} presents rich future work.
\begin{table}[t]
\caption{\userpmodel compared to baseline models reported in \citet{stelmakh2023gold} on a classification task constructed from user ratings. Easy Accuracy (EA) represents the ability to score relevant papers higher than irrelevant papers. Hard Accuracy (HA) on the other hand, represents the ability to score the more relevant paper of two relevant papers higher.}
\begin{tabular}{rcccc}
\textsc{RAPRatings} & EA & 95\% CI for EA & HA & 95\% CI for HA\\
\toprule
TPMS & 0.80 & [0.72, 0.87] & 0.62 & [0.54, 0.69]\\
TPMS$_{\text{FullText}}$ & 0.84 & [0.76, 0.91] & \textbf{0.64} & [0.56, 0.70]\\
ELMo & 0.70 & [0.62, 0.78] & 0.57 & [0.51, 0.63]\\
ACL & 0.78 & [0.69; 0.86] & 0.62 & [0.55; 0.68]\\
Specter & 0.85 & [0.76, 0.92] & 0.57 & [0.50, 0.63]\\
Specter+MFR & 0.88 & [0.81, 0.94] & 0.60 & [0.53; 0.66]\\
\midrule
LACE$_{|D_u|=20}$ & 0.87 & [0.79; 0.93] & 0.60 & [0.53; 0.67]\\
LACE$_{|D_u|=\text{All}}$ & \textbf{0.89} & [0.82; 0.95] & 0.59 & [0.52; 0.66]\\
\bottomrule
\label{tab-rapratings-clf}
\end{tabular}
\end{table}

\bibliographystyle{ACM-Reference-Format}
\balance
\bibliography{ed_exp-short}
\end{document}